**'SasCsvToolkit' - A versatile parallel `bag-of-tasks` job submission appli-
cation on heterogeneous and homogeneous platforms for Big Data Analytics
such as for Biomedical Informatics**

**Abhishek Narain Singh**
abhishek.narain@iitdalumni.com
Web: ABioTek www.tinyurl.com/abinarain

**Abstract**

**Background**: The need for big data analysis requires being able to process
large data which are being held fine-tuned for usage by corporates. It is only very
recently that the need for big data has caught attention for low budget corporate
groups and academia who typically do not have money and resources to buy ex-
pensive licenses of big data analysis platforms such as SAS. The corporates con-
tinue to work on SAS data format largely because of systemic organizational his-
tory and that the prior codes have been built on them. The data-providers continue
to thus provide data in SAS formats. Acute sudden need has arisen because of this
gap of data being in SAS format and the coders not having a SAS expertise or
training background as the economic and inertial forces acting of having shaped
these two class of people have been different.

**Method**: We analyze the differences and thus the need for SasCsvToolkit
which helps to generate a CSV file for a SAS format data so that the data scientist
can then make use of his skills in other tools that can process CSVs such as R,
SPSS, or even Microsoft Excel. At the same time, it also provides conversion of
CSV files to SAS format. Apart from this, a SAS database programmer always
struggles in finding the right method to do a database search, exact match, sub-
string match, except condition, filters, unique values, table joins and data mining
for which the toolbox also provides template scripts to modify and use from com-
mand line.

**Results**: The toolkit has been implemented on SLURM scheduler platform as a
`bag-of-tasks` algorithm for parallel and distributed workflow though serial ver-
sion has also been incorporated.

**Conclusion**: In the age of Big Data where there are way too many file formats
and software and analytics environment each having their own semantics to deal
with specific file types, SasCsvToolkit will find its functions very handy to a data
engineer.

ical informatics, R, Bioconductor, Toolkit

**Background**

The need for big data science keeps on increasing if not exponentially then def-
initely in a polynomial graph curvature as the industry has merged with techniques
in machine learning, Artificial Intelligence and High Performance Computing
(HPC) which in itself has grown remarkably, thereby having a direct symbiotic
growth effect for demand in big data science. Traditionally, the data science was practiced by and large by corporate firms who could pay hefty license fee for the tools that have tailor made solutions to a large number of issues which the open source movement do not offer. These tools such as SAS offered specialized procedures and features such as reading files without loading them completely into memory, unlike in open source tools like R which first loads the data as a data frame into memory before operating for any functionality check on them. SAS (pronounced "sass") once stood for "statistical analysis system", however now SAS covers more than just statistics as it engages machine learning and other complex computing also into its system ( https://www.sas.com/en_us/home.html). R is an open source language for statistical computing and graphics (https://www.sas.com/en_us/home.html). In this article, apart from enlightening the differences SAS and open source tools like R world, we also provide a toolkit that can convert a single or a set of SAS7BDAT file into CSV format to be used by R and other applications. At the same time, the SAS users might also be interested in doing some operations on a CSV format file to SAS format file, for which the tool also gives conversion script flexibility much in the same way of bag-of-tasks. The scripts can be executed either on a standalone serial mode, or in a parallel environment batch submission serial mode, or in a `bag-of-task` oriented parallel distributed mode.

**Need for SAS7BDAT data to CSV data format conversion & vice-versa Toolkit**

The corporate world, which could afford hefty license fee for tools such as SAS did not have to worry about big data challenges as much as it later became apparent of the open source world such as those using R. A lock in situation arose as all the IT organizational memory in terms of prior-codes, dependency prior-codes, know-how and any custom made training program for new hires were all in SAS. On the other hand, as the need for data science and big data analysis also became apparently important in the not so rich corporate, training, research and academic world alike, they grew their emphasis on R making it more and more popular. The gap between people wanting to work on SAS and people wanting to work on R thus simply kept increasing largely because of the inertia of certain skill sets and orientation acquired and thus the switching cost involved in terms of effort needed. While, the big memory requirements that a single R session would need for big files cannot be simply changed without some intelligent parallel programming, as it is in SAS, at least the data scientist can be in some comfort zone if the data provided to him is in Comma Separated Value (CSV) format rather than a SAS7BDAT format. Given the recent increase in the demand for R programmers despite the lacking features in R compared to SAS, due to the above reasoning I gave, there has been efforts by people to write modules or packages that can easily import the sas7bdat files to CSV files. These efforts have been very recent in only last 2-3 years, clearly emphasizing the need where R professionals are wanting to take over the roles which SAS professionals have had been having so far in corporate world. These additional packages such as haven by Wickham et. al. 2016 [1] or sas7bdat by Shotwell et. al. 2015 [2] thus became recently very popular as many companies have data in the order of several gigabytes in the sas7bdat format, and that can be easily dealt with an R or SPSS professional if they were simply converted to CSV format or at least loaded into data-frame without the need for CSV format conversion. The problem with these additional packages is that they fail to perform well on higher data sizes of capacity 100s of gigabytes to start with. More importantly, if there be need to read a section of the data, such that the parsing can be done in parallel for several different sections without the need for a high memory for the system such as not in the order of 100s of GB of RAM but few GBs of RAM, these packages will provide inherent deadlock situation. Even more importantly, any attempt to convert the read sas7bdat file in the form of data frame to a CSV file can be even more challenging in term of computational resources utilization of RAM, time and disk space simply because the CSV files will write the blank entries as 'NA' and not leave it blank without even a space, and mark every entry enclosed in a set of double quotes (""). The need for a tool that can facilitate the conversion of a large SAS7BDAT file to CSV format or even 100s of SAS7BDAT files to CSV format in an automated fashion is thus immense, given the time, cost and effort the data scientist can save this tool which the R packages does not currently offer at least efficiently for big files of sizes over 100 gigabytes to start with. At this point, I must revisit the data provider's business professionals. As the professionals in data collection and selling business are aware of the merits of big data high performance computing technologies, and are already locked in for their demands from corporate world for their profits, they tend to generate as much data whatsoever and in preferably SAS7BDAT format. As an example, in the health analytics sector, the Truven MarketScan database, an IBM enterprise, provides all of its data in SAS7BDAT format, and even if you purchase a 10 year commercial and medicare data that would be about 3.5 Tera-Bytes of SAS7BDAT format files to deal with. There have already been several biomedical publications citing Truven MarketScan database already, as an example of recent publications [6,7,8,9], as data science gets more popular even in the biomedical sector, given that it had already gained importance in big data bioinformatics and genomics sector. An R expert by no means and by no currently known freely available package will be able to even read the data unless you have a 1 Terabyte capacity of RAM on your supercomputer. The SasCsvToolkit finds its critical application to cater to the underpinning requirements of ever getting popular big data science world which is reaching out to every industry in being the game changer.

At the same time, a SAS expert would be having his comfort zone for data analytics in the SAS environment and would simply like to convert all CSV format data to SAS7BDAT format. Even for people who know both SAS and other data processing tools like SPSS, MATLAB, R, AWK, SED, PERL, know that some things can be done easier in one tool than the other and there can be situation that conversion to a SAS7BDAT format would be making more sense if the process to do downstream analysis is simpler in SAS systems. One such example could be to simply use SQL procedure in SAS in case SQL server is missing from the envi-ronment and the way to do inner joins and other table query might be difficult in AWK for instance or even using the JOIN command on Linux for the need for sorting and data spacing issue to be taken care of. Thus, a parallel distributed tool for converting large and/or many such CSV files to SAS7BDAT format would be an acute necessity.

The conversion of CSV to SAS7BDAT format also accelerates the benefits as the toolkit also comprises of sample template scripts for a database table search, exact match, substring match, except condition, filters, unique values, table joins and data mining which a SAS database programmer always struggles in finding the right method to develop the codes for it.

**SasCsvToolkit and its Usage**

The SasCsvToolkit is written in shell script and has SAS 9.4, Perl and Linux OS dependency. The scripts can be used to submit to a job submission scheduler such as the portable batch system, PBS, *Nitzberg* et. al. [3] or via SLURM, Jette et. al. [4], for which automation pipelines with parameters that can be adjusted in the file has been provided in the toolkit. The toolkit also comprises of automated SASSQL routine scripts for conducting basic search, string pattern match and inner joins that are a usual requirement for any database work.

A 'README' file is incorporated that well documents the usage syntax for the tool. The parallel jobs are implemented using Perl scripts and shell scripts combination and does not require an MPI, CUDA, OpenMP or Posix threads at all on top of the SLURM scheduler. The environment variables and SAS system command line variables have been used in a clever way to write files and communicate within script workloads.

Before using the SasCsvToolkit, it is critical to change the directory in the .sas extension files to where the library for SAS files are or are planned to be generated. A sub-directory called TrialDir is provided where the results will be generated. Sample small CSV files are also provided in this TrialDir for an execution to try out hands for CSV to SAS format conversion. To keep things easily identifiable and traceable for any error the log files are being generated for each sub-tasks or SAS files and the log file bears the name of the sub-task SAS file and its corresponding submission Id. The file sampleSasYbdatFileNames.txt has been provided as an example file which bears the name of the files in each line without the .sas7bdat extension. It is advised that the user should read the README file for more clarity about various kinds of usage options that exists for CSV to SAS format conversion or vice-versa in serial command line, batch submission or parallel mode.

**Materials and Resources**

The computing resources used were that of high performance facility, HPC. SLURM (simple Linux utility for resource manager) scheduler was already installed on the HPC, and SAS module 9.4 was loaded before execution. Perl v5.10.1 was preinstalled and there was no need for its module to be loaded. The .SAS7BDAT files of size about 3.5 TB was purchased from IBM Healthcare Tru-ven enterprise. The coding was done completely by the first author apart from algorithm design.

**Implementation**

The implementation of the toolkit is based on the fact that various commands in the SAS environment can be separated for its input using a delimiter such as a pipe '|'. Where needed the values that are fed at the command line can be simply de-limited for various options for input file names, expected output file names and the chain of pipeline procedure using the '|' pipe character. Each argument is then processed in a systematic fashion in the code as command line arguments are picked up. The SasCsvToolkit was deployed for IBM Healthcare Truven MarketScan Database primarily, though it can be used for other datasets as well, as many functionalities can be used straightaway by simply changing or adjusting the variable names. Many of the functions have the option of specifying the column variable of interest such as a in a relational database. SLURM functionalities to ensure bag-of-task algorithm has been implemented, where a bag-of-task mode of job submission ensures that there is no bottleneck which would otherwise have been present in case a barrier synchronization would have been used. A bag-of-task method loads all the independent units of tasks on the scheduler, and the scheduler then distributes the tasks based on whichever computing core gets free. The bag-of-tasks implementation in this case has achieved great speed up by ensuring maximum utilization of available computing cores in parallel, as we get to the numeric comparison as shown in figure 1, which is titled as ParaSASCSV to reflect the parallel nature of the toolkit.

The `bag-of-tasks` algorithm is used typically for a centrally automated robust parallel execution of highly independent sub-tasks. More importantly, if each of the sub-tasks takes decent amount of time for completion, and the bandwidth for read/write operations is not a limiting factor due to the cores being connected by infiniband network and also in physical proximity; the inter-processor communication overheads are relatively miniscule. In such a situation, deploying a `bag-of-tasks` algorithm makes immediate sense before venturing into further exploration of complicated parallel algorithmic strategies. Benoit et. al. in 2010 [5], gives details of how a `bag-of-tasks` algorithm is specifically useful for a heterogeneous computing environment. A bag-of-tasks in a naïve sense would mean that if I have a bag which has objects in it that needs to be processed by my workers, I first distribute 1 work to each worker, and then me as a master or a scheduler keep an eye on which of my worker gets his work completed. As soon as any of the workers finishes the given task, I take out another task from my bag and give it to the now available worker. The objects in my bag are in a state of queue to get processed by my workers who are active on the object job in hand. I need not wait for all of my workers to become free before I distribute my next set of jobs to them, such as is the case by setting a barrier synchronization, if there is a need for an all worker dependency. This real life example holds the same for a computing world where one computing core on a node can be the master or head computer and he uses a scheduling tool via a 'bag-of-tasks' approach to distribute new jobs to worker processing cores. The working nodes more or less acts as a slave core in this kind of distributed computing as it does not respond back to the head node to say it does not want any more job to be done. The beauty of independence of this methodology exploitation is that the computing worker nodes can be as heterogeneous in terms of their processor power, and as long as the memory requirement for the codes are met, the heterogeneity of the slave cores will not affect the smooth execution of the parallel and distributed job.

The algorithm can be broadly stated to comprise of following key steps:

- Distribute data files to various processors in a queue system where each job waits for its turn
- Perform independent functions on data files in parallel on the basis of dependency-analysis
- Head node gives data to slave for task operation wherever concurrency occurs

The following steps which are otherwise needed in a highly-interdependent parallel job was of no significance:

- No need to create barrier for synchronization
- No need to concatenate the generated intermediate result before switching to the next task
- No need to combine the final output and remove redundancy since the work is coarse-granulated and exhaustively complete

**Results**

The SasCsvToolkit was able to process data at a remarkable speedup of near linear values. The IBM Healthcare database which was about 3.5 TB in size comprising of about 160 files, and the automated pipeline successfully converted the data to desired CSV format in less than 48 hours using on average 10 running cores at a time in the serial SLURM job submission pipeline using the toolkit. 99% of the files took less than 4 hours for the completion with a RAM memory demand of about 32 GB with an average consumption of 3 hours per file, while the remaining 1% of the files took between 5-6 hours for conversion. If the serial version code in the SasCsvToolkit would have been used for conversion, the total time for completion could have been $160*3 = 480$ hours = 20 days. Thankfully, the tool also implemented a `bag-of-tasks` based algorithm for independent job submission, using which the whole completion of work took less than 2 days. The estimated speedup is about 10 times using the modest number of cores. It is by no surprise that the speedup is about the same as the number of active cores during execution which gives a good performance metric for the parallel version of the tool for a high efficiency close to 1.

*Estimated speedup = time for serial job / time for parallel job = 20 days / 2 days = 10 times.*

*Parallel efficiency = time for serial job/ (time for parallel job \* number of Processors) = 20 / (2\*10) = 1*

This linear speedup efficiency is achieved solely due to the fact that individual jobs are time consuming with each taking over 3 hours, and that there is minimum inter-processor communication needed due to the independence nature of each large chunks of granularity of the jobs. Figure 1, gives a graphical visualization for the comparison. With the increased speed-up comes high amount of risk reduction simply due to the fact that each processor now handles data which can fit well into its memory and does not generate any garbage value in addition to output files, as is the case for serial implementation for big data tasks with not adequate memory capacity of the core processors. This will be called a risk mitigation strategy in addition to increased efficiency, because many a times, the processor generates results without even informing the user that a lot of raw data was simply not read, as is indicative by the generation of garbage values which might not be produced at all. The SasCsvToolkit, nevertheless can be used even in serial mode and the benefit of the tool is not only the parallel implementation but also added several routine functionalities for which the user can simply follow the protocol syntax to do operations on relational databases.

**Discussion**

The SAS7BDAT tool thus thot only saves programming time needed for a big data scientist to convert the SAS7BDAT files to CSV format, but its built in parallel capability saves tremendous amount of time, given that multi-&-many core architectures in computing systems have become the norm these days, and so a software must make good use of the available resources. The toolkit works nevertheless for even a single core access machine, should there be scarcity of resources.

A detailed performance evaluation such as by varying data sizes and quality of data, varying the number of nodes and cores, and varying the quality of machine is well possible it has been done in many of previous publication on Para-Seqs, Singh et. al. in 2017 [10] is well possible to be done for SasCsvToolkit as well. However, given that the performance for scalability was almost 1:1 for 10 cores parallel job submission, which woule be considered very good, any further evaluation would not make much of a sense for an added value. Contrary to my previous publication [10], where the parallel versions of codes were only written by me, the value of the work presented in this paper is not just the parallel version of codes but also the serial versions of codes written by me given the increasing need for such as tool due to the gaps that has been formed between SAS and R/SPSS professionals, because of issues discussed earlier. The limitation of the current work is that the job distribution system used is that of SLURM and not the more popular PBS (portable batch submission) system. In future, such an implementation can also be made available.

**Conclusion**

The SasCsvToolkit proves to be a robust tool for a Linux / Unix environment for big data projects and its power and applicability increased many-folds by the wrapper Perl and Shell scripts that manages submitting multiple serial jobs to a SLURM job scheduler, as it saves an immense amount of time, cost and effort which would otherwise takes months to solve it, and thus helps the analyst get into the data analysis phase much quicker than without having this toolkit. The data science world in not only R, biomedical informatics, and big-data science world would be benefited, but also the SAS users themselves as by the use of this tool they can now explore operations that were perhaps not available in SAS but available in R, SPSS and other CSV processing utilities.

**Requirements**

Operating system(s): LINUX or its variants
Programming language: PERL
Other requirements: MPI, SAS module 9.4
License: MIT License
Any restrictions to use by non-academics: As per MIT License rules

**Abbreviations**

SAS - Statistical Analytics System, CSV - Comma Separated Values, MPI - Message Passing Interface, SLURM - Simple Linux Utility Resource Manager, HPC - High Performance Computing, PBS - portable batch submission

**Declaration Section**

**Ethical Approval and Consent to participate**

Not applicable

**Consent for publication**

Not applicable

**Availability of supporting data**

All supporting data are provided in this article.

**Competing interests**

None to be declared.

**Authors' contributions**

All work, idea generation, coding, analysis of results and writing the paper has been done by the first author.

**Acknowledgements**

The author acknowledges the makers of SAS software system.

**Authors' information**

This will be provided once the article is accepted. Until then it is hidden to keep anonymity.

**Figure Legend**

Figure 1: A rough comparison of serial versus parallel mode of operation for SAS7BDAT toolkit where number of processors used for parallel mode is 10.

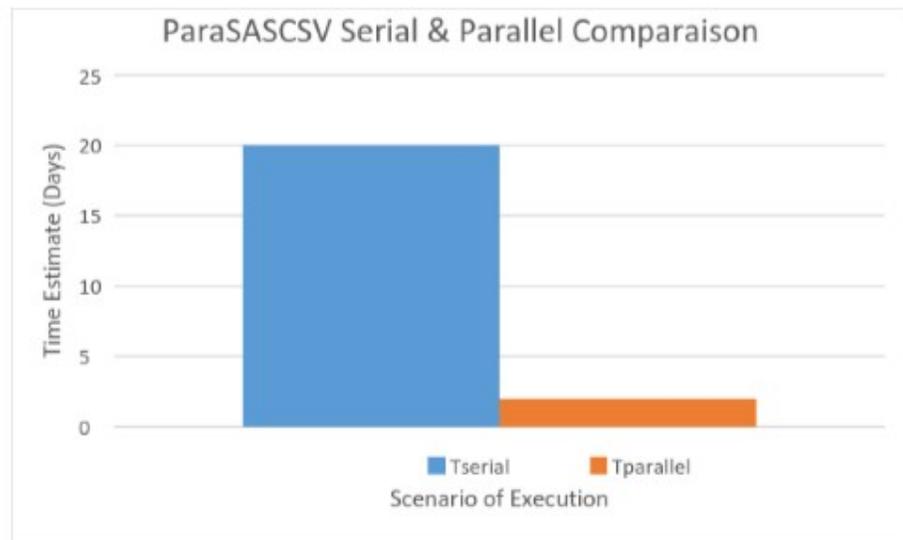

Figure 1